\documentclass[aps,nofootinbib,superscriptaddress,reprint,preprintnumbers]{revtex4-1}
\pdfoutput=1
\usepackage[nolist,nohyperlinks]{acronym}
\usepackage{graphicx,hyperref,braket,orcidlink,algorithm2e,amssymb,amsmath,dsfont,subfigure,slashed,bm,orcidlink}
\usepackage[capitalise,nameinlink]{cleveref}
\usepackage{lineno}
\graphicspath{{figs/}}
\hypersetup{
	unicode=true,
	pdftoolbar=true,
	pdfmenubar=true,
	pdffitwindow=false,
	pdfstartview={FitH},
	pdftitle={},
	pdfauthor={}, 
	pdfsubject={},
	pdfcreator={},
	pdfproducer={}, 
	pdfkeywords={},
	pdfnewwindow=true,
	colorlinks=true,
	linkcolor=blue,
	citecolor=olive,
	filecolor=magenta,
	urlcolor=cyan
}

\newcommand{\ie}{\textit{i.e.}\xspace}
\newcommand{\eg}{\textit{e.g.}\xspace}
\newcommand{\nabu}{\texttt{nabu}\xspace}
\newcommand{\EOS}{\texttt{EOS}\xspace}
\newcommand{\spey}{\texttt{Spey}\xspace}
\newcommand{\pvalue}{$p$-value\xspace}
\newcommand{\pvalues}{$p$-values\xspace}

\begin{document}


\title{Communicating Likelihoods with Normalising Flows}
\author{Jack Y. Araz\orcidlink{0000-0001-8721-8042}}
\email{jack.araz@stonybrook.edu}
\affiliation{Center for Nuclear Theory, Department of Physics and Astronomy, Stony Brook University, 11794, NY, Stony Brook, USA}
\author{Anja Beck\orcidlink{0000-0003-4872-1213}}
\email{anja.beck@cern.ch}
\affiliation{Massachusetts Institute of Technology, Cambridge, 02139, MA, USA}
\author{M\'eril Reboud\orcidlink{0000-0001-6033-3606}}
\email{merilreboud@gmail.com}
\affiliation{Université Paris-Saclay, CNRS/IN2P3, IJCLab, 91405 Orsay, France}
\author{Michael Spannowsky\orcidlink{0000-0002-8362-0576}}
\email{michael.spannowsky@durham.ac.uk}
\affiliation{Institute for Particle Physics Phenomenology and Department of Physics, Durham University, Durham DH1 3LE, UK}
\author{Danny van Dyk\orcidlink{0000-0002-7668-810X}}
\email{danny.van.dyk@gmail.com}
\affiliation{Institute for Particle Physics Phenomenology and Department of Physics, Durham University, Durham DH1 3LE, UK}

\date{\today}
\preprint{IPPP/25/07}

\begin{abstract}
    We present a machine-learning-based workflow to model an unbinned likelihood from its samples.
    A key advancement over existing approaches is the validation of the learned likelihood using rigorous statistical tests of the joint distribution, such as the Kolmogorov-Smirnov test of the joint distribution.
    Our method enables the reliable communication of experimental and phenomenological likelihoods for subsequent analyses.
    We demonstrate its effectiveness through three case studies in high-energy physics.
    To support broader adoption, we provide an open-source reference implementation, \texttt{nabu}.
\end{abstract}

\maketitle

\acrodef{LM}{Likelihood Model}
\newcommand{\LM}{\ac{LM}\xspace}
\acrodef{ML}{Machine Learning}
\newcommand{\ML}{\ac{ML}\xspace}
\acrodef{MLP}{Multilayer Perceptron}
\newcommand{\MLP}{\ac{MLP}\xspace} 
\acrodef{NF}{Normalizing Flow}
\newcommand{\NF}{\ac{NF}\xspace}
\acrodef{KS}{Kolmogorov-Smirnov}
\newcommand{\KS}{\ac{KS}\xspace}
\acrodef{PDF}{Probability Density Function}
\newcommand{\PDF}{\ac{PDF}\xspace}
\acrodef{CDF}{Cummulative Density Function}
\newcommand{\CDF}{\ac{CDF}\xspace}
\acrodef{MAF}{Masked Autoregressive Flow}
\newcommand{\MAF}{\ac{MAF}\xspace}
\acrodef{RQS}{Rational Quadratic Spline}
\newcommand{\RQS}{\ac{RQS}\xspace}

\section{Introduction}

The reinterpretability of experimental results is a cornerstone of progress in high-energy physics, where the sheer expense and effort required to generate data preclude the luxury of rerunning experiments.
In collider experiments, for instance, the complexity of the data and the sophisticated workflows needed for its analysis mean that sharing results in a way that enables reinterpretation is both a scientific imperative and a technical challenge~\cite{Alguero:2022gwm, Araz:2021akd, Araz:2020lnp, Conte:2018vmg, Dumont:2014tja, Conte:2014zja, Conte:2012fm, Araz:2023axv, Bierlich:2019rhm, Buckley:2019stt, Buckley:2010ar, Alguero:2021dig, Alguero:2020grj, Ambrogi:2018ujg}.
Traditional approaches, such as sharing either the raw experimental data or likelihoods, present significant limitations. 
Raw data sharing is resource-intensive and raises concerns about data management and accessibility~\cite{opendata, Aidala:2023dai}.
Providing full likelihoods is computationally expensive and often impractical due to large file sizes and the high number of experimental nuisance parameters ~\cite{Heinrich:2021gyp, Cranmer:2012sba}.
Simplified likelihoods mitigate these issues but introduce assumptions that can overly constrain their usability~\cite{Araz:2023bwx}. 

We suggest to use \acp{LM} instead.
A \LM is a mathematical representation that quantifies the probability of observing a specific data set given underlying parameters. In high-energy physics, likelihood models are essential for interpreting experimental results and comparing theoretical predictions against observed data. These models strive to encapsulate the complete statistical information of an experiment, making them a cornerstone for parameter estimation \cite{Flacher:2008zq}, hypothesis testing \cite{Englert:2015hrx,Englert:2017aqb}, and model comparison \cite{Drees:2013wra,Dercks:2016npn,GAMBIT:2017yxo,Amrith:2018yfb}. However, constructing and utilizing these models can be computationally challenging, mainly when dealing with high-dimensional parameter spaces or complex workflows that include detector effects and systematic uncertainties.

\ML has emerged as a transformative tool in high-energy physics in recent years. \ML-based approaches offer potent solutions for complex, high-dimensional problems arising in detector simulation, event reconstruction, and parameter inference. Among these, \acp{NF} stand out as a promising avenue for constructing surrogate models of likelihoods. By learning the underlying probability distribution of experimental data, \acp{NF} can provide flexible and accurate approximations of the likelihood function that are computationally efficient to evaluate. These surrogate models enable data sharing in a compact, interpretable format and facilitate downstream tasks such as inference and model comparison. Previous work in this direction (see, for example, Refs.~\cite{Coccaro:2019lgs, Reyes-Gonzalez:2023oei, Beck:2023xou}) has demonstrated the potential of \acp{NF} to encapsulate the essential features of high-dimensional distributions while remaining computationally tractable.

A parallel challenge in high-energy physics lies in the computational cost of the simulation pipeline. Monte Carlo simulations, augmented by parton showering and detector simulations, are indispensable for modelling experiment interactions. However, their expense often limits their use, especially in scenarios requiring iterative evaluation, such as Bayesian inference or optimization of theoretical models. Differentiable programming workflows~\cite{Heinrich:2022qlq,Smith:2023ssh,Adelmann:2022ozp,Heimel:2022wyj,Kagan:2023gxz,Kofler:2024efb,Vigl:2024lat,Kagan:2023gxz,Heimel:2023ngj,Heimel:2024wph}, which integrate gradients throughout the simulation pipeline, offer a potential solution. These workflows allow efficient navigation of the parameter space but are currently limited by the monumental effort required to adapt existing simulations to a fully differentiable framework.

Our work bridges these gaps by employing \acp{NF} to build surrogate models that approximate both the likelihood function and the simulation workflow, aligning well with simulation-based inference~\cite{Cranmer:2019eaq, Brehmer:2019xox, Brehmer:2018eca, Brehmer:2018kdj, Mastandrea:2024irf, Dax:2023ozk}. This dual application accelerates parameter inference and extends the surrogate models' utility to data generation and pseudo-experimental analysis tasks. By designing the surrogate models to capture the essential features of the experimental data while being accessible to a wide range of researchers, we ensure that our approach supports the community’s broader goals of reproducibility and reinterpretability~\cite{Araz:2023mda}.

The following sections outline a framework for \NF-based surrogate models to represent experimental likelihoods. We provide practical methodologies for their training and validation and demonstrate their application to various high-energy physics use cases. Additionally, we discuss how these surrogate models can serve as a stepping stone towards a fully differentiable simulation pipeline, enabling efficient, iterative exploration of theoretical and experimental parameter spaces. By intentionally designing our framework to be user-friendly and broadly applicable, we ensure that it can be adopted across the community for diverse reinterpretation and inference tasks.

\section{Methods and step-by-step recipe}

As part of our workflow, we pursue the following objectives:
\begin{itemize}
    \item constructing a \LM from pre-existing samples;
    \item testing the compatibility of this \LM with respect to the samples; and
    \item storing the \LM for future use, e.g. the evaluation of the model density or generation of additional samples.
\end{itemize}
A Python implementation of our workflow is available through the open source software \nabu~\cite{NABU:Software},\footnote{%
Named after the Assyrian patron god of scribes and wisdom.
}
which is built on \texttt{flowjax}~\cite{flowjax}, \texttt{equinox}~\cite{equinox}, and \texttt{jax}~\cite{jax}.
The full documentation of \nabu~will be presented in a future publication.\\

\noindent
\textbf{Preparations}~In many cases, a ``standardization'' of the samples (\ie transformation of the samples to a parameter space with standard properties) is useful before training.
We support such standardization, \eg, through transforming each variable in the $D$-dimensional data set to new variables with zero mean and unit variance.
In high-energy physics, Dalitz analyses yield a special case of non-Cartesian bounded parameter space, as shown in one of the
examples below. For such cases, we provide the means to transform the samples to a Cartesian parameter space.
After the optional ``standardization'' step, the workflow splits the available data set of size $N$ into a training set ($N_\text{tr} \simeq 72\% \times N$ by default),
a validation set ($N_\text{v} \simeq 8\% \times N$ by default), and a testing set ($N_\text{test} \simeq 20\%\times N$ by default).
\\

\noindent
\textbf{Model}~We train a \LM $\tilde{L}$ on the training data set, denoted $\vec{\vartheta}_1$ to $\vec{\vartheta}_{N_\text{tr}}$,
with each sample \mbox{$\vec{\vartheta}_i \in T \equiv \mathbb{R}^D$}.
Following the convention of Ref.~\cite{Beck:2023xou}, we refer to the vector space $T$ as the \emph{target space}.
If the \LM $\tilde{L}$ is trained sufficiently and successfully, it can serve as a proxy for the true likelihood $L(\vec{\vartheta})$.

A \NF $f(\cdot)$ is constructed to transform the training samples from the target space into the base space $B$,
\begin{equation}
    f(\vec\vartheta_i) = \vec\beta_i \in B \equiv \mathbb{R}^D\,.
\end{equation}
Our default choice of \acp{NF} employs \MAF~\cite{Papamakarios:2017tec} alternated with
configurable permutations. Our implementation~\cite{NABU:Software} also supports further types of \acp{NF},
including e.g. non-volume preserving transformations (RealNVP)~\cite{Dinh:2016pgf}.
We keep the number of flow layers configurable.
Each flow layer is associated with a \ac{MLP} to learn the functions used in the underlying bijections.
The width, depth, and choice of activation functions of these \acp{MLP} are configurable. As a transformation function we employ either a simple affine transformation or \RQS~\cite{durkan2019neuralsplineflows} where the parameters of the transformation has been set via a \ac{MLP}.
The trainable parameter set is comprised of the union of the parameters of all \acp{MLP}.\\

The \LM is trained by minimising a loss function $\ell_\text{tr}$ with respect to these parameters.
This training is performed on a configurable number of batches of equal sample size $N_\text{batch}$. For each batch,
we define its loss $\ell$ as the mean value of the unbinned standard-normal likelihood of the samples in base space,
\begin{equation}
    \ell_\text{tr} = -\frac{1}{N_\text{batch}}\sum_{i=1}^{N_\text{batch}} \ln \left[\mathcal{N}^D(\vec\beta_i \,|\, \vec{\mu} = \vec{0}, \Sigma = \mathds{1})\right]\,.
\end{equation}
We minimise this loss for each batch, using the ``ADAM''~\cite{Kingma:2014vow} minimiser by default, with an adjustable learning rate.
A training epoch is completed once every batch has been used for minimisation.
At this point, we record the mean of the loss value across all training batches.\\

We repeat the loss computation using the validation set, yielding $\ell_\text{v}$.
The training procedure stops either after a configurable number of epochs have passed or if the validation loss $\ell_\text{v}$ stops decreasing for
a configurable number of epochs.
\\

\noindent

\textbf{Testing}~After completion of the training, the base-space samples should follow a standard normal distribution
if the \LM has adequately learned the features inherent to the training sample set.
We test if this is true by using a suitable test statistic.
A training outcome is rejected if its \pvalue is smaller than a pre-defined threshold (default: $3\%$).

Our primary choice of test static is the unbinned \KS~\cite{kstest} test on the \PDF for the square of the two-norm of the transformed samples $\vec\beta_i$.
For a $D$-dimensional standard normal \PDF of the samples, the \PDF of the squared two-norm is a $\chi^2$ distribution with $D$ degrees of freedom.
As a consequence of having a known joint \PDF, we can efficiently perform the \KS test on our training and validation sets after every epoch,
if so requested by the user.
Our approach has benefits compared to other works in the literature~\cite{Reyes-Gonzalez:2023oei}.
First, our approach utilizes a single \emph{known} distribution that tests the joint distribution rather than each marginal \PDF.
Second, the known \CDF increases the performance of the \KS test and removes the Monte-Carlo uncertainty for the \pvalue,
which arises from the limited size of the generated testing sets in two-sample tests, \eg for unknown distributions or distributions
without an evaluable \CDF.

Our secondary choice of the test statistic is primarily intended for visualization: we split the two-norm samples $||\vec\beta_i||_2^2$ into $K$ bins of equal probability content.
For a large number of samples, $N / K \gtrsim 100$,
the number of samples in each bin follows a normal distribution with $\mu = N / K$ and $\sigma = \sqrt{N / K}$.
We use this expected mean and standard deviation to perform a binned $\chi^2$ test on combining the first $K-1$ bins.
\\

\noindent
\textbf{Storage}~We provide the means to store a \LM in a single file
representing the \NF and all coordinate transformations applied to the sample set before training.
This approach facilitates third parties' dissemination and usage of the \LM.

\section{Concrete Examples}

\begin{figure}
    \subfigure[
        \ ATLAS data~\cite{atlas_collaboration_2024_11507450}
        from Ref.~\cite{ATLAS:2024xxl}.
        \label{fig:results:ATLAS}
    ]{%
        \includegraphics[width=.35\textwidth]{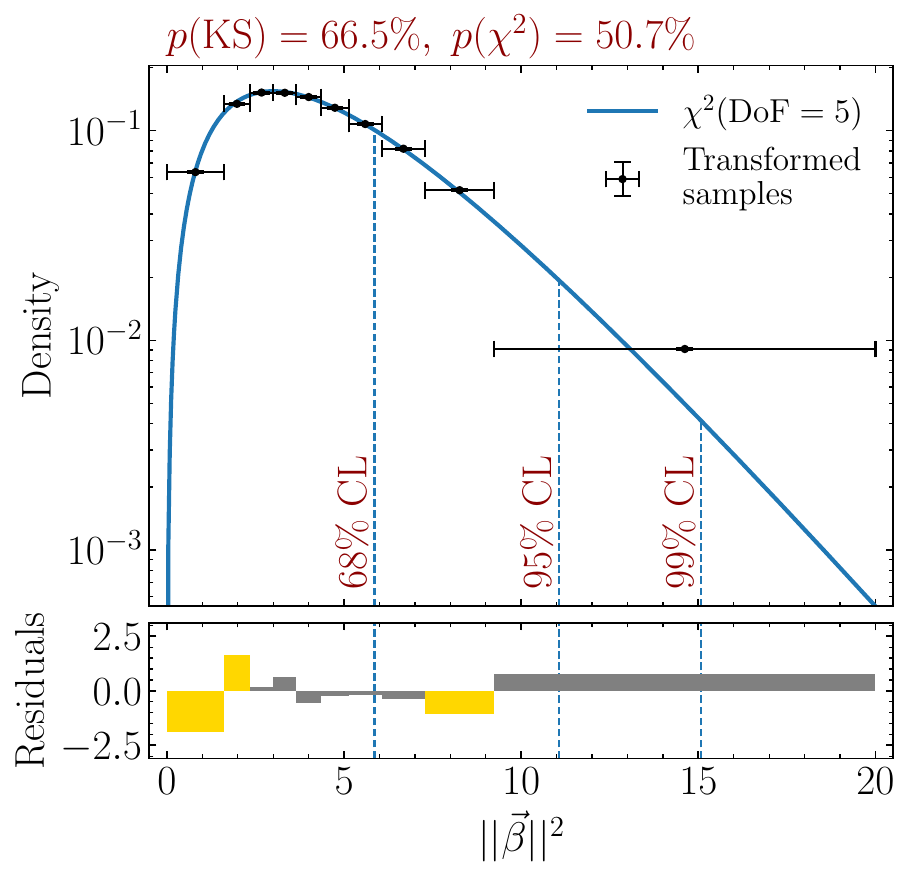}
    }
    \\
    \subfigure[
        \ LHCb Dalitz plot data from Ref.~\cite{LHCb:2020pxc}.
        \label{fig:results:LHCb}
    ]{%
        \includegraphics[width=.35\textwidth]{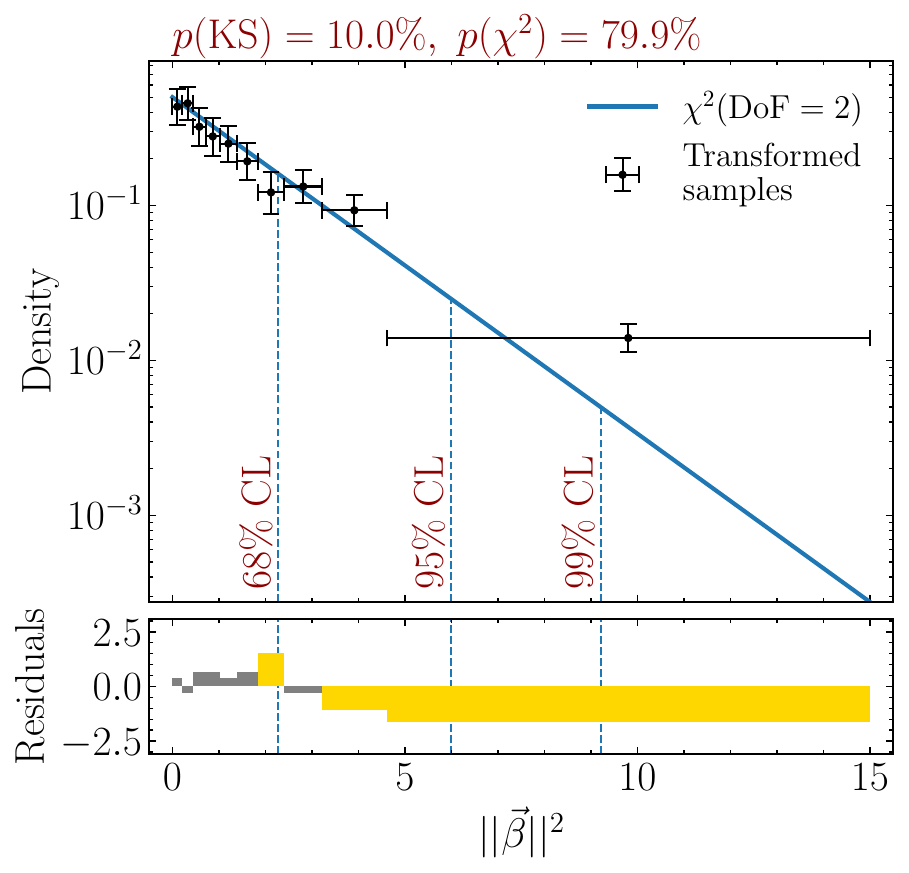}
    }
    \\
    \subfigure[
        \ WET $b\to u\ell^-\bar\nu$ samples~\cite{EOS-DATA-2023-01} from Ref.~\cite{Leljak:2023gna}.
        \label{fig:results:WET}
    ]{%
        \includegraphics[width=.35\textwidth]{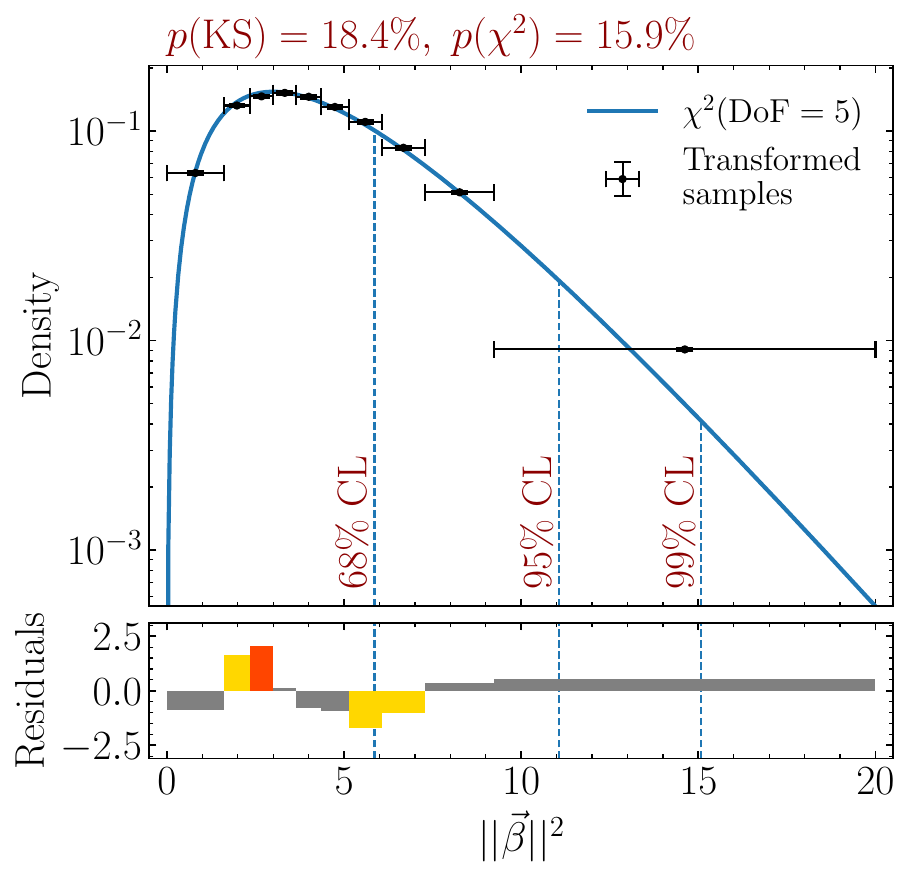}
    }
    \\
    \caption{
        Summary plots for the three training examples.
        We show the \pvalue corresponding to the \KS and binned $\chi^2$ tests.
        The ``Density'' part of each plot shows the 10 bins in $||\vec\beta||^2$ and the expected $\chi^2$ \PDF as a blue curve.
        Each bin is expected to contain 10\% of the testing set for a perfect model.
        In the ``Residuals'' part of each plot, the grey, gold, and red coloured bins indicate deviations of less than $1\sigma$, within $[1,2)\sigma$, and more than $2\sigma$, respectively.
    }\label{fig:results}
\end{figure}

We illustrate the performance of our workflow using real-world examples. Our selection of examples
covers experimental and phenomenological results with a variety of different dimensionalities and complexities
of distribution.
\Cref{fig:results} contains a summary plot for each example.
The upper part of each plot shows the distribution of the two-norms of the transformed samples after successful training.
Overlaid is the corresponding expected $\chi^2$ distribution.
The calculated \pvalues are presented at the top, and the residuals part of each plot illustrates the similarity between the empirical and expected distributions. The significance of each residual is shown using colour coding, with
yellow bins indicating $1\sigma$ to $2\sigma$ fluctuations and red bins indicating a fluctuation beyond $2\sigma$.

In \hyperref[fig:results:ATLAS]{example (a)}, we train a \LM on $pp \to Z(\to \mu\mu) + \text{jet}$ events~\cite{atlas_collaboration_2024_11507450} published by the ATLAS experiment~\cite{ATLAS:2024xxl}.
The original data set features $24$ dimensions. However, in the underlying $2\to 3$-body process,
most physical information is expected to be encoded in only five independent observables.
We choose $p_T^{\mu\mu}$, $y^{\mu\mu}$, $p_T^{\mu_1}$, $\Delta \eta_{12}^2=\eta_{\mu1}^2 - \eta_{\mu2}^2$, $\Delta \phi_{12}^2=\phi_{\mu1}^2 - \phi_{\mu2}^2$ as representative observables for this physical information.
The data set has been standardized by scaling and shifting all parameters to the range $[0,1]$.
This is followed by applying a logarithmic transformation to the modified $p_T^{\mu\mu}$ and $p_T^{\mu_1}$ distributions and rescaling them to $[0,1]$ once more.
This approach mitigates challenges in learning distributions that extend to very large values.

An eight-layer \NF is employed for training, utilizing \MAF bijections with \RQS transformations.
Each \MAF layer contains a single-layer \MLP with 512 hidden units, while the \RQS transformation is
configured with 12 knots within the $[0,1]$ support.
The initial learning rate is set to $1\%$ and decays with a half-life of 25 epochs until reaching a minimal learning rate of $10^{-6}$. 
The \NF is trained for a maximum of 600 epochs, with early stopping applied if the negative log probability of the validation set does not improve for more than 50 epochs. Testing the \LM yields a \KS \pvalue of $66.5\%$.
The outcome of this example is illustrated in \cref{fig:results:ATLAS}.
The resulting model file uses $\sim 4\text{MB}$ of storage space, while the compressed samples use $\sim 14\text{MB}$.

In \hyperref[fig:results:LHCb]{example (b)}, we train an \LM on efficiency corrected events of the decay $B^+\to D^+ D^- K^+$
as published by the LHCb experiment~\cite{LHCb:2020pxc}.
The events have been extracted from the supplementary material. Each event corresponds to a 2-tuple of the squared $D^- K^+$ mass and the squared $D^+D^-$ mass.
Before training, we manually map the support of the Dalitz plot to the space $(-\infty,+\infty)^2$.
For this two-dimensional example, we require 8 flow layers of type \RQS,
and an \MLP with two hidden layers \& 64 neurons per layer to learn the likelihood.
After testing, our \LM yields acceptable \pvalues of $10\%$ and $79.9\%$ for the \KS test and the binned $\chi^2$ test, respectively.
The current LHCb measurement does not yet
provide sufficient detail to highlight resonance bands in $m_{D^+D^-}^2$ as expected from the presence of
broad charmonium resonances above the open-charm threshold. On the other hand, we are content that the
\LM does not ``hallucinate'' non-existing structures.
This example is, therefore, a good illustration of the stability of our choice of \LM.
The resulting model file uses $\sim 300\text{kB}$ of storage space, while the samples use $\sim 14\text{kB}$.

In \hyperref[fig:results:WET]{example (c)}, we train an \LM on the posterior samples of the Weak Effective Theory parameters
in the $ub\ell\nu$ sector~\cite{EOS-DATA-2023-01}, obtained from a fit to exclusive
$\bar{B}\to \lbrace \pi,\rho,\omega\rbrace \ell^-\bar\nu$ decays~\cite{Leljak:2023gna}.
This data set contains 5-dimensional samples following a multimodal distribution.
The underlying data is the same as used for the pilot study presented in Ref.~\cite{Beck:2023xou}.
In contrast to earlier efforts, our present approach is \emph{successful} in learning the full five-dimensional
distribution, as indicated by its acceptable \pvalues of $18.4\%$ and $15.9\%$ for the \KS test and the binned $\chi^2$ test, respectively.
As a consequence, the results of Ref.~\cite{Leljak:2023gna} can now be used in subsequent
BSM model studies or SMEFT studies, without repeating the original analysis. This represents a substantial
reduction of the computation costs, given the large number of $50$ hadronic nuisance parameters that
are irrelevant to any BSM implications.
The resulting model file uses $\sim 90\text{kB}$ of storage space, while the compressed samples use $\sim 5\text{MB}$.

\section{Conclusion and Outlook}

Analysis preservation and reinterpretation of experimental and phenomenological analyses is a significant hurdle
to progress in the field of high-energy physics.
Here, we have presented a workflow to model unbinned likelihoods for this purpose.
When constructing likelihood models, we make use of established techniques (such as normalizing flows)
and pieces of software that emerged in the context of machine-learning applications.
We combine these techniques with robust statistical tests that ensure the likelihood model is accurate.
Using three concrete real-world examples from experiment and phenomenology,
we have illustrated that our proposed workflow can model likelihoods with complicated non-Gaussian features and strong correlations.
Except for our two-dimensional example, the storage size of our models is much smaller than the size of the training samples.
This confirms an advantage of our proposed workflow for data preservation and storage.
\\

We will continue developing the \nabu software~\cite{NABU:Software}, expanding its capabilities with additional types of normalising flows and a broader range of known transformations.
Future applications of \nabu include integration with the \EOS software for flavour physics~\cite{EOSAuthors:2021xpv}, enabling precise posterior constraints on Weak Effective Theory parameters inferred from flavour-changing processes~\cite{Leljak:2023gna,Bolognani:2024cmr,Meiser:2024zea}.
Additionally, \nabu will interface with \spey~\cite{Araz:2023bwx} to accelerate analysis reinterpretation by constructing a comprehensive catalogue of likelihood models from publicly available experimental data.
Looking ahead, \nabu is designed for extendability beyond normalising flows, including advanced machine learning techniques such as diffusion models, which we aim to explore in future work.

\section*{Acknowledgement}

We thank Peter Stangl for his helpful comments on the manuscript.
MS and DvD acknowledge support by the UK Science and Technology Facilities Council (STFC) through grant ST/X003167/1.
DvD further acknowledges UK STFC support through grant ST/V003941/1.
AB wants to acknowledge the support of MIT's Department of Physics.

\newpage
\bibliography{references.bib}

\end{document}